\documentclass[letterpaper]{article}

\usepackage{natbib,alifeconf}  
\usepackage{booktabs}
\usepackage{bm}
\usepackage{amsmath}    
\usepackage{amssymb}    
\usepackage{url,hyperref,cleveref}
\usepackage{color}
\usepackage{tikz}
\usepackage{soul}

\def\pr#1{\left( #1\right)}

\def\ang#1{\left\langle #1\right\rangle}

\title{Nonequilibrium Thermodynamics of Associative Memory\\ Continuous-Time Recurrent Neural Networks}

\author{
    Miguel Aguilera$^{1,3,*}$,
    Daniele De Martino$^{2,3}$,
    Ivan Garashchuk$^1$ \and
    Dmitry Sinelshchikov$^{2,3}$
    \\
    \mbox{}\\
    $^1$BCAM -- Basque Center for Applied Mathematics, 48009 Bilbao, Spain \\
    $^2$Biofisika Institute, 48940 Leioa, Bizkaia, Spain \\
    $^3$IKERBASQUE, Basque Foundation for Science, 48009 Bilbao, Spain \\
    $^*$maguilera@bcamath.org
} 

\begin{document}

\maketitle

\begin{abstract}
    Continuous-Time Recurrent Neural Networks (CTRNNs) have been widely used for their capacity to model complex temporal behaviour. However, their internal dynamics often remain difficult to interpret. In this paper, we propose a new class of CTRNNs based on Hopfield-like associative memories with asymmetric couplings. This model combines the expressive power of  associative memories with a tractable mathematical formalism to characterize fluctuations in nonequilibrium dynamics. We show that this mathematical description allows us to directly compute the evolution of its macroscopic observables (the encoded features), as well as the instantaneous entropy and entropy dissipation of the system, thereby offering a bridge between dynamical systems descriptions of low-dimensional observables and the statistical mechanics of large nonequilibrium networks. Our results suggest that these nonequilibrium associative CTRNNs can serve as more interpretable models for complex sequence-encoding networks.
\end{abstract}

Submission type: \textbf{Full Paper}\\


\section{Introduction}
Continuous-Time Recurrent Neural Networks (CTRNNs) are a common tool to model adaptive behaviour in neural systems and artificial agents \citep{beer1997dynamics, beer1996toward,di2008not,izquierdo2008associative,yamashita2008emergence}. Their ability to display complex continuous-time dynamics has made them particularly suitable for tasks involving the interaction of active perception and motor control. However, one of the persistent challenges in employing CTRNNs is the lack of interpretability in their internal representations and emergent dynamics.

In parallel, `associative memories' or Nakano-Amari-Hopfield networks \citep{nakano1971learning, amari1972learning, hopfield1982neural}---usually referred to as Hopfield networks---provide a mathematically elegant framework for associative memory with interpretable attractor dynamics. While usually considered in equilibrium setups with symmetric network couplings, these models can be extended to display nonequilibrium dynamics, showing that asymmetric couplings allow the storage and retrieval of sequences \citep{amari1972learning, sompolinsky1986temporal,during1998phase}.
These asymmetric associative networks constitute a general class of non-reciprocal interacting systems \citep{fruchart2021non}, capable of exhibiting out-of-equilibrium phase transitions and, most notably, collective nonlinear oscillations recently reported in nonequilibrium spin models \citep{de2019feedback,sinelshchikov2023emergence,guislain2023nonequilibrium,guislain2025far}. Moreover, they can be used to describe the behaviour of modern AI systems like attention layers in transformer models \citep{krotov2023new,poc2024dynamical}.
These developments suggest a rich yet underutilized potential for using associative memories to engineer interpretable networks encoding sequential information patterns. 

In this work, we propose a new class of CTRNNs grounded in associative memories dynamics with asymmetric couplings in the thermodynamic limit. Our approach defines a stochastic, time-continuous update rule whose nonequilibrium nature enables sequence encoding and temporal generalization while maintaining tractable analytical tools. 
We use existing models and theory on asymmetric associative memories and extend it to describe stochastic fluctuations of the model's mean-field dynamics. 
We demonstrate that, through a change of variables, the model can be reformulated into low-dimensional CTRNN-like model, with explicit interpretations of memory and pattern overlap. Furthermore, we derive new expression capturing the entropy production and entropy change, characterizing the nonequilibrium thermodynamics of the system.

\section{Continuous-Time Recurrent Neural Networks}

CTRNNs are widely used to model the time evolution of neural activations in artificial agents. Each neuron's activity is governed by a leaky integrator with recurrent input and nonlinear feedback. The dynamics of a CTRNN are typically written as:
\begin{equation}
\bm{\tau} \odot \dot{\bm{y}} = -\bm{y} + \bm{W} \, \sigma(\bm{\theta} + \bm{y}) + \bm{I},
\end{equation}
where $\bm{y} \in \mathbb{R}^N$ is the vector of neural activations, $\bm{\tau} \in \mathbb{R}^N$ is the vector of membrane time constants, $\bm{W} \in \mathbb{R}^{N \times N}$ is the recurrent weight matrix, $\bm{\theta} \in \mathbb{R}^N$ is a bias vector, $\bm{I} \in \mathbb{R}^N$ is an input vector, and $\sigma(x) = (1 + e^{-x})^{-1}$ is a sigmoid activation function. The element-wise multiplication $\odot$ represents each neuron's integration time scale.

An equivalent and formulation of the model above places the synaptic weights inside the nonlinearity and changes the nonlinear function as:
\begin{equation}
\bm{\tau} \odot \dot{\bm{z}} = -\bm{z} + \tanh(\bm{\theta} + \bm{W} \bm{z} + \bm{I}),
\label{eq:CTRNN}
\end{equation}
which is mathematically equivalent under a change of variables and redefinition of parameters (under a change in the activation function $\sigma(x) = [1+\tanh(x)]/2$).

\section{Asymmetric associative memories as nonequilibrium dynamical systems}

Classical associative memories are defined by simpler binary neurons $\bm x \in \{\pm 1\}^N$, coupled by symmetric connectivity matrices $\bm J$ ($J_{ij} = J_{ji}$), which ensure the existence of a Lyapunov (`energy') function and guarantee convergence to fixed-point attractors under asynchronous dynamics. This is equivalent to the Ising model in statistical physics. 
While analytically convenient, this symmetry condition imposes strong constraints that are neither biologically realistic nor dynamically rich once the system approaches a steady state. In particular, symmetric networks are restricted to gradient dynamics and lack the ability to generate sequential behaviour, such as limit cycles, strange attractors and chaos.

To lift these constraints, we consider asymmetric connectivity matrices, $J_{ij} \neq J_{ji}$, which break detailed balance and drive the network into nonequilibrium regimes. A natural way to define such couplings is through a low-rank structure \citep[see][]{coolen2001statistical}:
\begin{equation}
J_{ij} = \frac{1}{N} Q(\bm{\xi}_i, \bm{\xi}_j),
\end{equation}
where $\bm{\xi}_i \in \mathbb{R}^M$ (typically $\bm{\xi}_i  \in \{\pm 1\}^M$) encodes a set of pattern features for neuron $i$, and $Q$ is a non-symmetric kernel determining effective interactions between pattern components. A simple choice for kernel $Q$ is a quadratic relation
\begin{equation}
    Q(\bm{\xi}_i, \bm{\xi}_j) = \sum_{ab} A_{ab} \xi_i^a \xi_j^b,
\end{equation}
where $\bm A$ is a non-symmetric $M$x$M$ matrix with $A_{ab} \neq A_{ba}, a,b = 1, \ldots, M$ encoding pattern interactions.

Unlike their symmetric counterparts, these networks cannot be described by equilibrium statistical mechanics. Instead, they require tools from nonequilibrium physics, such as generating functional analysis, to characterize dynamical observables and quantify their entropy production \citep{coolen2001statistical,aguilera2023nonequilibrium}

\subsection{Stochastic dynamics and mean-field Langevin approximation}

The kinetic Ising system consissts of binary states  $x_{i} \in \{-1, +1\}$, representing the state of spin $i \in \{1,\dots,N\}$. This can be interpeted as the state of a neuron in a firing or silent state. Each spin receives an input from other spins $x_j$ through $J_{ij}$. 
The transition rate for a single spin flip is given by the Glauber rule
\begin{align}
    w(\bm x) = \frac{1}{2}(1+x_{i}\tanh(\beta h_{i})),
    \label{eq:glauber}
\end{align}
with a local field
\begin{align}
    h_{i} = \sum_a \Theta_a \xi_i^a  + \frac{1}{N} \sum_{ab,j} A_{ab} \xi_i^a \xi_j^b x_{j}.
\end{align}
Here, $\bm \Theta = (\Theta_1,\ldots, \Theta_m)$ is added as an external field promoting the emergence of a particular encoded pattern.

The spin distribution can be described by a stochastic master equation, where spins update asynchronously \citep{coolen2001statistical}.
\begin{align}
    \dot p(\bm x) = \sum_i \left( w(x_i\,|\,\bm x^{[i]})p(\bm x^{[i]}) - w(-x_i\,|\,\bm x)p(\bm x) \right).
    \label{eq:master-equation}
\end{align}
Here the spin-flipped configuration $\bm x^{[i]}$ is obtained by inverting the sign of spin $x_i$.

We define the macroscopic order parameters as
\begin{align}
    m^a(\bm x) = \sum_i \xi_i^a x_i,
\end{align}
and describe the probability of a given overlap configuration $\bm m$ as
\begin{align}
    P(\bm m) = \sum_{\bm x} p(\bm x) \delta_{\bm m(\bm x), \bm m},
\end{align}
with $\delta_{\bm m(\bm x), \bm m} = \prod_a \delta_{m^a(\bm x), m^a}$. For large system sizes, applying the Kramers–Moyal expansion to the master equation we obtain
\begin{align}
    \dot P(\bm m)  =&  \sum_{i,\bm x} \delta_{\bm m(\bm x), \bm m} \pr{w(\bm x^{[i]})p(\bm x^{[i]}) - w(\bm x)p(\bm x)}
    \nonumber \\  \approx &   \nabla_{\bm m}  \bigg(\sum_{\bm x} \delta_{\bm m(\bm x), \bm m} \frac{2}{N} \sum_i \bm \xi_i s_i  w(\bm x)p(\bm x) \bigg)
    \nonumber \\ & + \nabla^2_{\bm m}  \bigg( \sum_{\bm x} \delta_{\bm m(\bm x), \bm m} \sum_i \frac{2}{N^2} \bm \xi_i^\top \bm \xi_i  w(\bm x)p(\bm x)\bigg)
    \nonumber \\  =&   \nabla_{\bm m}  \bigg( -\bm m + \frac{1}{N} \sum_i \bm \xi_i  \nu (\bm  \xi_i^\top)   P(\bm m) 
      \nonumber \\ & +   \frac{1}{N^2}\sum_i \bm \xi_i \bm \xi_i^\top \bigg(1- \mu (\bm  \xi_i^\top)  \, \nu (\bm  \xi_i^\top) \bigg)  \nabla_{\bm m}   P(\bm m)  \bigg) \label{eq:Kramers-Moyal}
\end{align}
with $\mu (\bm  \xi_i^\top) = \sum_{\bm x} x_i p(\bm x)\delta_{\bm m(\bm x), \bm m}$ and $\nu (\bm  \xi_i^\top) = \tanh(\beta \bm \xi_i^\top (\bm \Theta + \bm A \bm m))$. Note that, assuming equiprobable states conditioned to a value of $\bm m(\bm x)$, we can derive $\mu (\bm  \xi_i^\top) = \tanh(\beta \bm \xi_i^\top \bm \theta)$ for a parameter $\bm \theta$ such that $\bm m(\bm x) = \sum_i \bm \xi_i \tanh(\beta \bm \xi_i^\top \bm \theta)$. This relation corresponds to the solution of a variational problem that maximizes the entropy subject to fixed mean overlap, and can be obtained by maximizing the Lagrangian $ L(\bm \theta) = N\bm\theta^\top \bm m - \sum_i \ln(2\cosh(\beta \bm \xi_i^\top \bm \theta))$, 
where $\bm \theta$ plays the role of Lagrange multipliers enforcing the overlap constraint.
Algorithmically, the values of $\bm \theta$ are easy to obtain, given $\bm m(\bm x)$, e.g., by the Newton-Raphson method. 

The result in Eq. \eqref{eq:Kramers-Moyal} takes the form of  a Fokker–Planck equation for $P(\bm m)$. Note that this extends the result obtained in \citep{coolen2001statistical} by including stochastic diffusion term that account for fluctuations in systems of finite size. Eq. \eqref{eq:Kramers-Moyal}  corresponds to the  Langevin dynamics:
\begin{align}
    \dot{\bm m} = -\bm m + \frac{1}{N} \sum_i \bm \xi_i \tanh( \beta \bm \xi_i^\top (\bm \Theta + \bm A \bm m) ) + \bm \eta,\label{eq:mean-field_raw}
\end{align}
where $\bm \eta$ is a Gaussian white noise term with the covariance
\begin{align}
    \bm \Gamma = \frac{1}{N^2} \sum_i \bigg( \bm \xi_i \bm \xi_i ^\top  - \bm \xi_i \bm \xi_i ^\top \mu(\bm \xi_i) \nu(\bm \xi_i) \bigg).
\end{align}
Note that, in the limit of large $N$, fluctuations vanish. 

The structure of Equation~\eqref{eq:mean-field_raw} closely resembles that of a CTRNN, where $\bm m$ plays the role of a low-dimensional neural state (e.g. a neural field). In some particular cases, we can recover the form of a regular CTRNN. For example, let $\{\mathcal{I}_a\}_{a=1}^M$ be a partition of the $N$ spins, with
$\mathcal{I}_a \subset \{1, \dots, N\}, \quad \mathcal{I}_a \cap \mathcal{I}_b = \emptyset \text{ for } a \ne b$, and $\xi_i^a = 
1 \text{ if } i \in \mathcal{I}_a$ and $\xi_i^a = 0$ otherwise. In this case, in the limit $N\to\infty$, Eq.~\eqref{eq:mean-field_raw} exactly recovers \eqref{eq:CTRNN}. 
More generally, \eqref{eq:mean-field_raw} represents a class of neural networks in which the  interaction term involves a matrix  $\bm A$ of encoded patterns $\bm \xi$. These encoded patterns generate nontrivial interactions in the network (e.g., competition between expressed patterns when they cannot be simultaneously activated).

Another distinction from standard CTRNNs lies in the stochastic component $\bm \eta$, which introduces state-dependent noise with a covariance matrix $\bm \Gamma$. This noise arises from the finite-size fluctuations in the underlying spin dynamics and is modulated by both the current mean-field state $\bm m$ and the structure of the patterns via $\bm \xi$. As a result, neurons  interact through a shared latent structure encoded in the patterns, making $\bm \xi$ the effective mediator of connectivity and memory structure. This formulation enables a compact, interpretable representation of memory-guided dynamics while capturing both deterministic and stochastic aspects of neural computation.

In practice, tracking all values of $\bm \xi$ is computationally difficult and involves an exponentially growing number of combinations. In order to overcome this problem, we will follow here the standard approach in statistical mechanics of disordered systems, tracing over the possible patterns $\bm{\xi}$ treated as `quenched' stochastic variables \citep{amit1985spin,mezard1987spin}, where \eqref{eq:mean-field_raw} can be simplified under different assumptions. 
In the simplest case, assuming $\bm \xi$ are uniform independent random variables $\xi_i^a = \pm 1$ (which is the case for encoding binary memories), the Fokker plank equation results in.
\begin{align}
    \dot{\bm m} =& -\bm m + \frac{1}{2^M} \bm S \tanh(\beta \bm S ^\top (\bm \Theta + \bm A \bm m)) + \bm \eta, 
    \label{eq:mean-field}
    \\
    \bm \Gamma =&  \,\, \frac{1}{N^2} \pr{\bm I  
   -\frac{1}{2^M} \bm S \pr{\mu( \bm S^\top ) \odot \nu(\bm S^\top )}\bm S^\top }.
\end{align}
where
$\bm S$ is a $M$x$2^M$ matrix, where columns collect all possible binary pattern configurations $\bm \xi_i \in \{-1,1\}^M$.
In the symmetric case, this scheme is known to work in the undersaturated regime $M \ll N$, which is the focus of this article.
In comparison with the standard CTRNN model, we observe that the matrix S introduces a new type of interactions, related to the interaction between the different expressed patterns.

\subsection{Thermodynamics of nonequilibrium associative memories}

To characterize the informational structure of the system under a fixed mean-field configuration $\bm m$, we assume that all microscopic spin states $\bm x$ compatible with $\bm m(\bm x) = \bm m$ are equiprobable. Under this assumption, the conditional distribution $p(\bm x \,|\, \bm m)$ is uniform over the manifold of configurations that yield the same overlap.

Decomposing the entropy of the system conditioned on $\bm m$
\begin{align}
    \mathcal{S} =& -\ang{\log p(\bm x)} =  -\ang{\log p(\bm m)} - \ang{\log p(\bm x|\bm m)}
\end{align}
In the large size limit, the term $-\ang{\log p(\bm m)}$ becomes negligible as fluctuations decrease with size. 
Under the assumption of equiprobable states for the same overlap value under the mean-field assumption $\bm m(\bm x) =  \frac{1}{N} \sum_i \bm \xi_i \tanh(\beta \bm \xi_i^\top \bm \theta)$ (where values of $\bm \theta$ have to be inferred numerically), we obtain
\begin{align}
    \mathcal{S} =& \ang{\log p(\bm x|\bm m))}
    \nonumber\\ =& \sum_i \ang{\ln (2\cosh(\beta \bm \xi_i^\top \bm \theta)) - N \bm \theta^\top \bm m },
\end{align}
which captures the number of microstates compatible with the macroscopic overlap $\bm m$. This expression quantifies the memory degeneracy of the network and sets the thermodynamic groundwork for calculating entropy production and fluctuation theorems in subsequent sections.
To quantify the system's nonequilibrium nature, we compute the entropy production. The entropy production rate measures the time-reversal asymmetry of the dynamics.

The entropy's time derivative yields the entropy production rate $\dot{\mathcal{S}} = \dot\Sigma - \Phi$, where $\Phi$ is the entropy flux to the environment, and $\dot\Sigma$ is the internal entropy production.
Using the standard expression from stochastic thermodynamics \citep{seifert2005entropy}, and noticing that the probability of the spin state after a flip attempt $x'_i$ is  $T(x'_i|\bm x) = \exp(\beta x'_i h_i)/(2\cosh(\beta h_{i}))$, the entropy production rate is calculated from the master equation \eqref{eq:master-equation} by:
\begin{align}
    \dot\Sigma =&  \sum_{\bm x, i}  T(x'_i|\bm x)p(\bm x)  \ln \frac{T(x'_i|\bm x)p(\bm x)}{T(x_i|\bm x')p(\bm x')}
    \nonumber\\ =& \dot{\mathcal{S}} + \sum_i \ang{ (x_i' - x_i) \beta h_i}
    \nonumber\\ =& \dot{\mathcal{S}} + \sum_i \ang{ (\tanh( \beta h_i ) - x_i) \beta h_i}
    \nonumber\\ =& \dot{\mathcal{S}} + N \ang{\beta  (\bm \Theta + \bm A \bm m)^\top  \dot{\bm m}}
\end{align}
In the second step, we assume that self-couplings are $J_{ii}=0$.
This expression highlights the time-asymmetry of nonequilibrium dynamics, as it quantifies the mismatch between forward and time-reversed overlap dynamics. When $\bm A$ is symmetric, $\dot\Sigma = 0$, recovering detailed balance.

Finally, the entropy change can be calculated as
\begin{align}
    \dot{\mathcal{S}} &=- \ang{\frac{d \log p(\bm x)}{d\bm m} \dot{\bm m}} 
    = -N \ang{ \bm \theta^\top \dot{\bm m}} 
\end{align}
where ${d\bm\theta}/{d\bm m}$ terms cancel out.

\subsection{Asymmetric associative memories as nonequilibrium CTRNNs}

The dynamics derived above reveal a direct relation between nonequilibrium associative memories and Continuous-Time Recurrent Neural Networks (CTRNNs). In particular, the macroscopic mean-field equation  \eqref{eq:mean-field} describes a leaky nonlinear integrator whose feedback term is shaped by asymmetric pattern couplings $\bm S$ through the matrix $\bm A$. In the absence of noise, this structure resembles the canonical CTRNN form. In addition, the pattern structure encoded in the weights of the system can give rise to complex dynamics. In some cases, pattern encoding enables designing networks whose dynamics can be interpreted from the encoding matrix $\bm A$. In others, $\bm A$ can lead to complex behaviours like chaos and complex thermodynamic signatures.

\subsubsection{Example: sequence encoding}
A simple way to test the network's sequence encoding capabilities is to impose a structured temporal ordering among patterns. For instance, setting the coupling matrix as $A_{ab} = \delta_{a, (b+1) \mathrm{mod} M} - k\, \delta_{a, (b+3) \mathrm{mod} M}$, creates a cyclic transition between successive patterns, where the parameter $k$ modulates the temporal frequency of the resulting oscillations.

\begin{center}
\begin{tikzpicture}[>=stealth, node distance=2cm]
  \node (dotsL) at (-1,0) {$\ldots$};
  \node (x0) at (0,0) {$\bm{\xi}^1$};
  \node (x1) at (1,0) {$\bm{\xi}^2$};
  \node (x2) at (2,0) {$\bm{\xi}^3$};
  \node (x3) at (3,0) {$\bm{\xi}^4$};
  \node (x4) at (4,0) {$\bm{\xi}^5$};
  \node (dotsR) at (5,0) {$\ldots$};

  \draw[->, thick] (dotsL) -- (x0);
  \draw[->, thick] (x0) -- (x1);
  \draw[->, thick] (x1) -- (x2);
  \draw[->, thick] (x2) -- (x3);
  \draw[->, thick] (x3) -- (x4);
  \draw[->, thick] (x4) -- (dotsR);

  \draw[-|, red, bend left=35] (x0) to (x3);
  \draw[-|, red, bend left=35] (x1) to (x4);
\end{tikzpicture}
\end{center}

\begin{figure}[h]
    \centering
    \begin{tikzpicture}
        \node[anchor=south west] (A) at (0,0) {\includegraphics[width=0.65\linewidth]{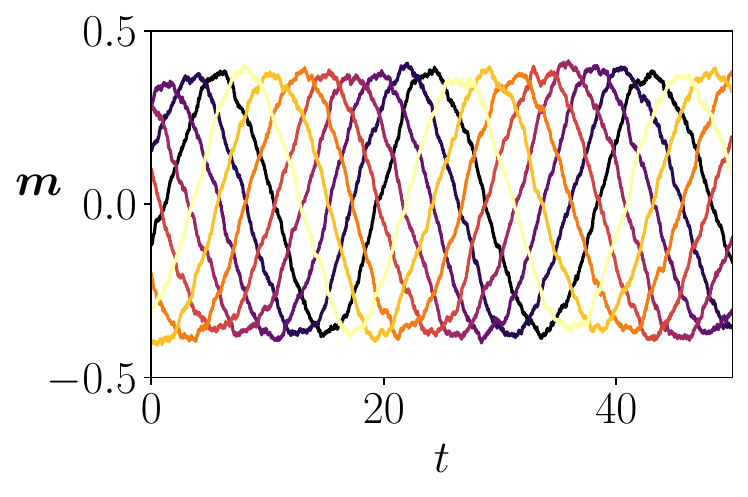}};
        \node[anchor=north west] at (A.north west) {\textbf{(a)}};
        \node[anchor=north west] (B) at ([yshift=-1mm]A.south west) {};
         \node[anchor=north west] (B1) at ([xshift=0]B.north west)
         {\includegraphics[width=0.65\linewidth]{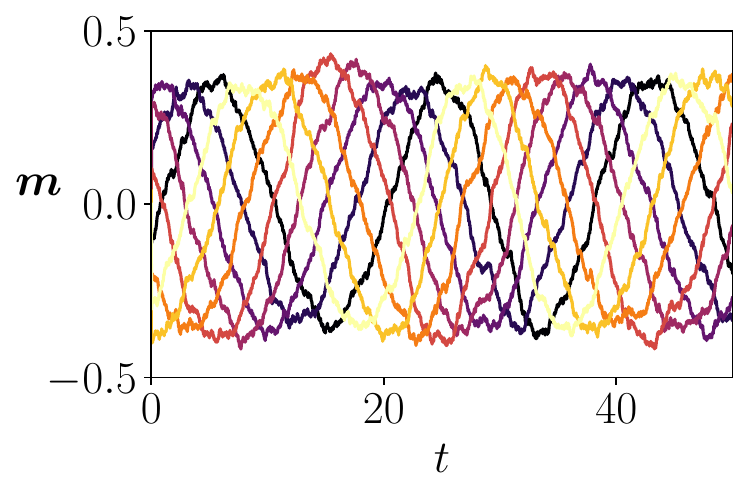}};
        \node[anchor=north west] at (B.north west) {\textbf{(b)}};
    \end{tikzpicture}
    \vspace{-4mm}
    \caption{Cyclic encoding of 8 patterns with $k=0.5$ with $N = 1000$ spins. a) Mean-field dynamics in  \eqref{eq:mean-field}. b) Monte Carlo simulation of the Glauber dynamics \eqref{eq:glauber}.}
    \label{fig:cycle}
\end{figure}
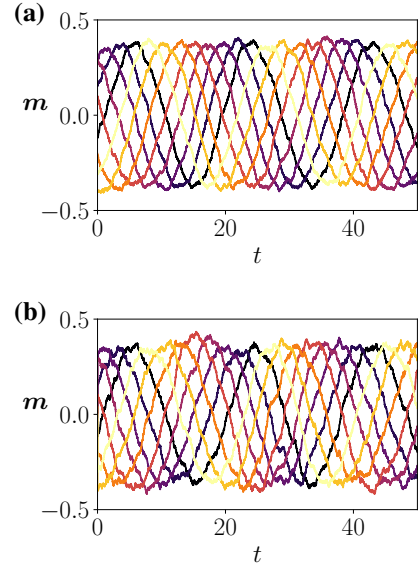

Figure~\ref{fig:cycle} illustrates this behaviour by comparing the mean-field dynamics (a) with a stochastic spin simulation of $N = 1000$ units (b) for a system  cyclically encoding 8 patterns. The qualitative agreement between both representations confirms that the low-dimensional mean-field equations accurately capture the emergent sequential dynamics of the full network.

This example demonstrates that nonequilibrium associative memories are capable of generating interpretable, structured sequences through low-rank asymmetric interactions. The dynamics are governed by a compact set of pattern-based variables and exhibit a robust cyclic activation of memory states. 

Such structured cyclic dynamics suggest potential applications in biologically inspired control architectures, such as central pattern generators  for rhythmic motor behaviors  \citep{marder2001central}. Future work may explore more complex forms of sequential encoding, including branching sequences, hierarchical memory structures, and context-dependent switching between stored trajectories.

\subsubsection{Example: chaotic system.}
To illustrate the model's dynamical richness, we present the chaotic regimes that we find in a four-dimensional system that describes the mean-field dynamics of the spin model \eqref{eq:mean-field}. To identify them, we integrate the system numerically and compute the full spectrum of the Lyapunov exponents using the Benettin algorithm \citep{Benettin1980}. To explore the high-dimensional parameter space (21 parameters) and localize the domains of chaos, we employ a genetic algorithm \citep{Mitchell1998} guided by a fitness function designed to promote chaos. 
We define fitness as the negative sum of the two largest Lyapunov exponents, $F = \lambda_1 + \lambda_2$. This objective 
aims to find chaos while reducing the occurrence of wide fitness plateaus at $\lambda_1=0$, corresponding to limit cycles and quasiperiodic attractors, which are widespread in the vicinity of the domains of chaotic dynamics. 

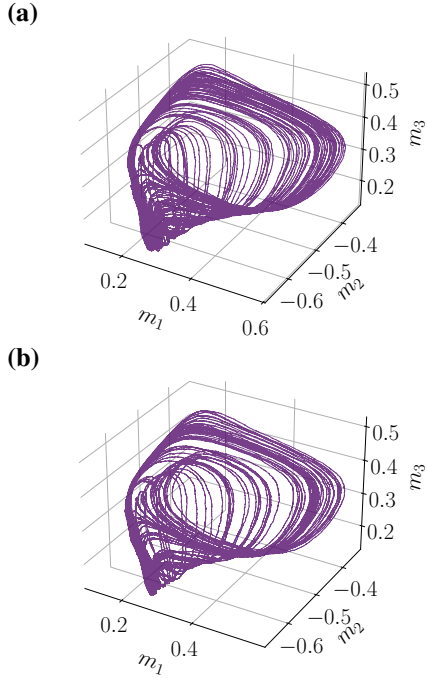
\begin{figure}
    \centering
    \begin{tikzpicture}
        \node[anchor=south west] (A) at (0,0) {\includegraphics[width=0.75\linewidth]{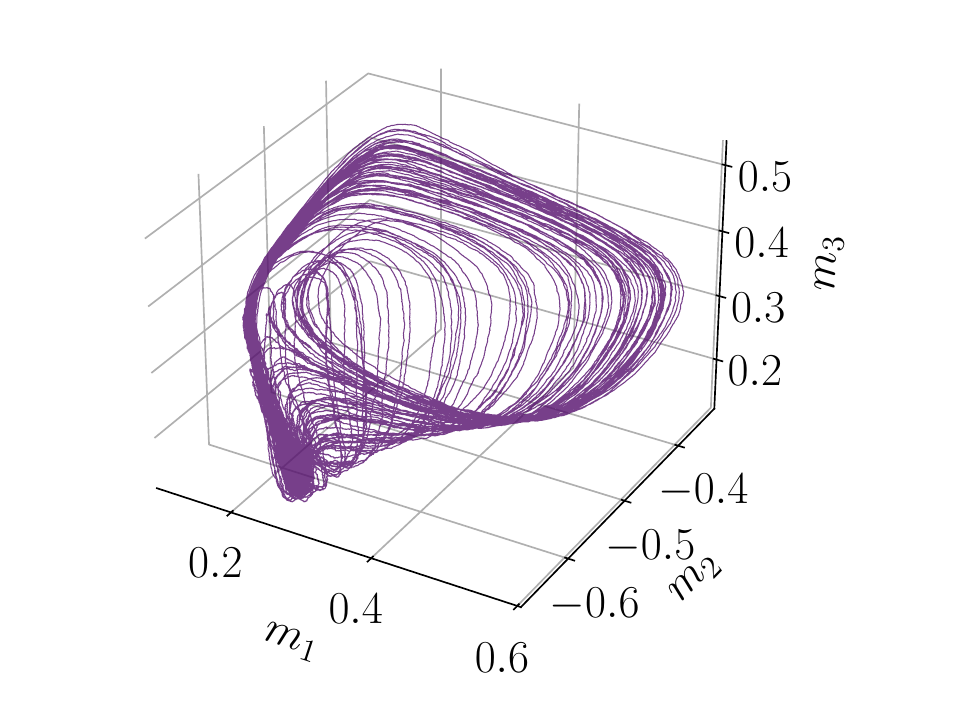}};
        \node[anchor=north west] at (A.north west) {\textbf{(a)}};
        \node[anchor=north west] (B) at ([yshift=4mm]A.south west) {};
         \node[anchor=north west] (B1) at ([xshift=0]B.north west)
         {\includegraphics[width=0.75\linewidth]{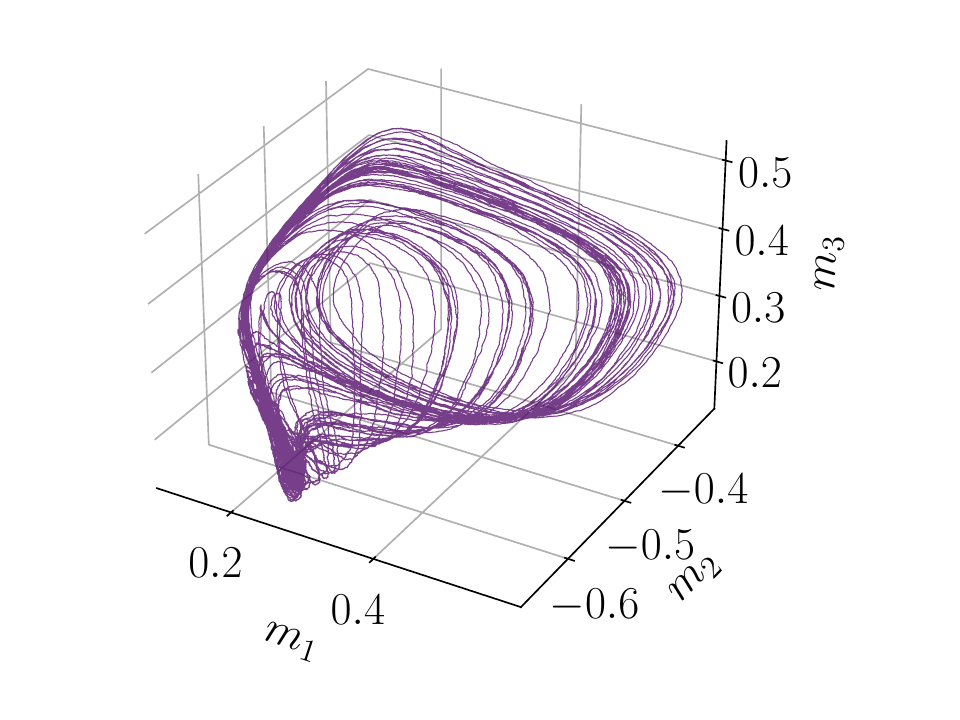}};
        \node[anchor=north west] at (B.north west) {\textbf{(b)}};
    \end{tikzpicture}
    \caption{Chaotic dynamics in a model with  4 patterns and $N = 100,000$ spin calculated for  \textbf{(a)} the mean-field dynamics  \eqref{eq:mean-field} and  \textbf{(b)} a Monte Carlo simulation of \eqref{eq:glauber}.}
    \label{fig:chaos}
\end{figure}

The algorithm uses a population size of 86\footnote{This specific number was chosen due to the number of CPU cores on a sever node.}, initially sampled from either the uniform or Gaussian distribution. 
We assume zero mean for the coupling parameters $A_{ij}$ and biases $\theta_i$, and a mean of $1$ for the inverse temperature $\beta$ (the final result was rescaled to have $\beta=1$). 
In each run of the genetic algorithm random initial conditions for the dynamical system are generated. We keep the initial conditions the same across the entire population and do not change during the optimization process to ensure convergence.

As the stop condition we set a maximum of generations to be 140 and the stall threshold of 30 generations without improvement (defined as a change in fitness of the best candidate in a population smaller than $2 \cdot 10^{-4}$). We apply a position-based crossover strategy with a crossover fraction of 0.8, and a Gaussian mutation operator with standard deviation decaying as $ 1/n$, where $n$ denotes the generation number. This setup allows efficient navigation of the parameter space while maintaining diversity and numerical stability.

We represent in Figure~\ref{fig:chaos} one of the possible configurations of parameters that produces chaotic dynamics, with $\beta=1$ and
\[
\bm{\Theta} = \left[\begin{smallmatrix}
-0.17096 \\
-0.45531 \\
2.79146 \\
-3.77300
\end{smallmatrix}\right], \,
\bm{A} = \left[\begin{smallmatrix}
3.7386 & 0.9059 & 7.2209 & 1.8722 \\
-1.7720 & 3.8092 & -0.0728 & -0.0080 \\
-3.9072 & 8.1931 & 5.8763 & -8.6256 \\
0.3361 & -0.0071 & 3.5843 & -0.4835
\end{smallmatrix}\right].
\]
The system is simulated by both integrating the mean-field equations \eqref{eq:mean-field} and running the stochastic Glauber dynamics  \eqref{eq:glauber} for a network of $N = 100,000$ spins. The displayed trajectory shows the complex temporal evolution of the pattern overlap vector $\bm{m}$ in a regime where multiple attractors and irregular transitions are present due to asymmetric couplings.
Chaos was found before in small CTRNNs with three nodes \citep{beer1995dynamics}, though this required the combination of different temporal integration variables $\bm \tau$ in Eq.~\eqref{eq:CTRNN}.

\begin{figure}
    \centering
    \begin{tikzpicture}
        \def\xshift{6mm}
        \node[anchor=south west] (A) at (\xshift,0) 
            {\includegraphics[width=0.8\linewidth]{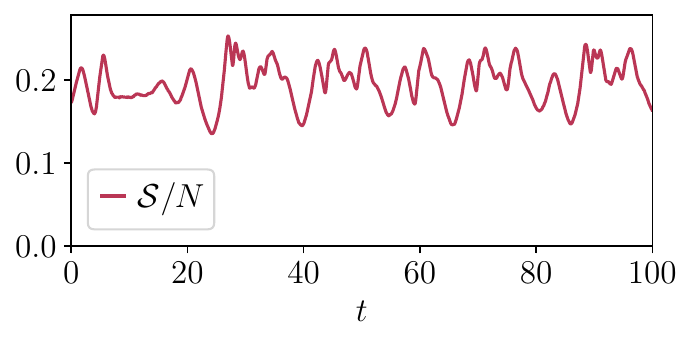}};
        \node[anchor=north west] at ([xshift=-2mm,yshift=2mm]A.north west) 
            {\textbf{(a)}};
    
        \node[anchor=north west] (B) at (A.south west) 
            {\includegraphics[width=0.8\linewidth]{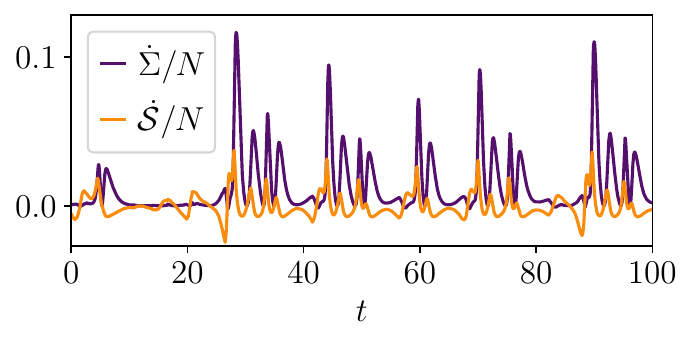}};
        \node[anchor=north west] at ([xshift=-2mm,yshift=2mm]B.north west) 
            {\textbf{(b)}};
    \end{tikzpicture}
    \caption{Time evolution of thermodynamic quantities in the chaotic regime. The plot shows the instantaneous entropy $\mathcal{S}$, its time derivative $\dot{\mathcal{S}}$, and the entropy production rate $\dot{\Sigma}$ computed from the mean-field trajectory of the chaotic system in Figure~\ref{fig:chaos}.}
    \label{fig:EP-time}
\end{figure}

\subsubsection{Nonequilibrium chaotic dynamics.}
To further characterize the nonequilibrium nature of the chaotic regime, we compute the instantaneous entropy $\mathcal{S}$, its time derivative $\dot{\mathcal{S}}$, and the entropy production rate $\dot{\Sigma}$ along the mean-field trajectory shown in Figure~\ref{fig:chaos}. These quantities are calculated using the analytical expressions above. The results, shown in Figure~\ref{fig:EP-time}, confirm that the chaotic system persistently generates entropy due to time-reversal asymmetry, as indicated by sustained positive values of the entropy production rate $\dot{\Sigma}$. Interestingly, we observe rare transient dips where $\dot{\Sigma} < 0$, corresponding to local fluctuations that momentarily violate the second law---consistent with predictions from fluctuation theorems. While both $\dot{\mathcal{S}}$ and $\dot{\Sigma}$ fluctuate significantly, the time-averaged entropy change vanishes, $\langle \dot{\mathcal{S}} \rangle = 0$, indicating that the system remains in a nonequilibrium steady state with stationary entropy. These thermodynamic quantities provide a principled and interpretable view of the system’s chaotic dynamics.

\begin{figure}[h]
    \centering
    \begin{tikzpicture}
        \def\xshift{6mm}
        \node[anchor=south west] (A) at (\xshift,0) 
            {\includegraphics[width=0.9\linewidth]{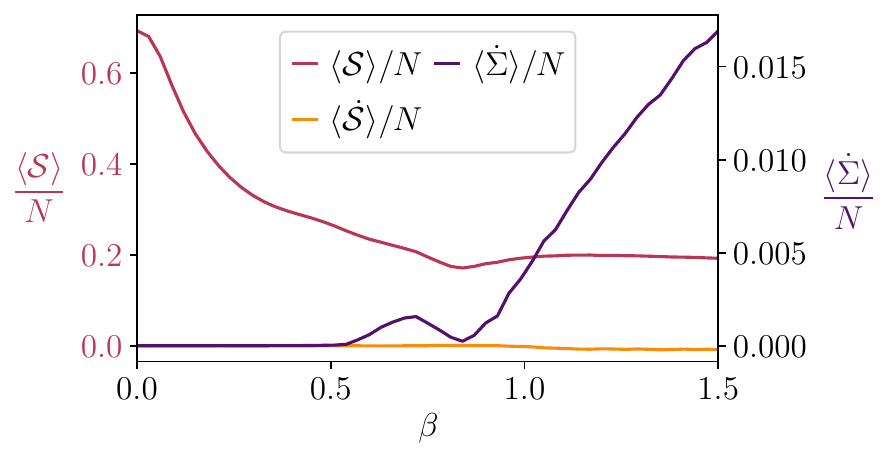}};
        \node[anchor=north west] at ([xshift=-1mm,yshift=2mm]A.north west) 
            {\textbf{(a)}};
    
        \node[anchor=north west] (B) at ([xshift=-0mm]A.south west) 
            {\includegraphics[width=0.75\linewidth]{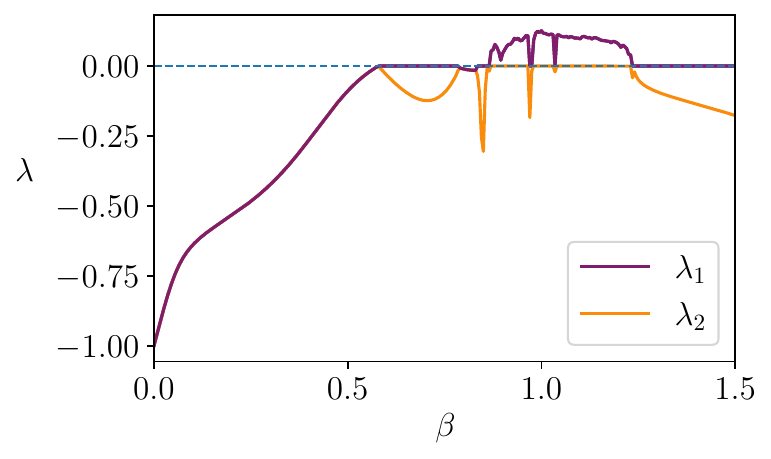}};
        \node[anchor=north west] at ([xshift=-1mm,yshift=2mm]B.north west) 
            {\textbf{(b)}};
    \end{tikzpicture}
    \caption{ (a) Average of thermodynamic quantities in the chaotic regime. The plot shows the instantaneous entropy $\mathcal{S}$, its time derivative $\dot{\mathcal{S}}$, and the entropy production rate $\dot{\Sigma}$ computed from the mean-field trajectory of the chaotic system at different values of $\beta$. \textbf{(b)} Largest two Lyapunov exponents for each value of $\beta$, reflecting periodic (largest exponent equal to zero) and chaotic (positive largest exponent) regions.}
    \label{fig:EP-lyapunov}
\end{figure}

Figure~\ref{fig:EP-lyapunov}(a) displays the average thermodynamic quantities as a function of inverse temperature $\beta$. We observe that the entropy production rate $\langle \dot{\Sigma} \rangle$ becomes appreciable as $\beta$ increases, reaching a maximum in the chaotic regime, while the average entropy change $\langle \dot{\mathcal{S}} \rangle$ remains zero, as expected in a steady state. The average entropy $\langle \mathcal{S} \rangle$ tends to reduce with $\beta$ (except at the onset of chaos, see below), reflecting changes in microscopic state accessibility as the system becomes more constrained.

Figure~\ref{fig:EP-lyapunov}(b) plots the largest two Lyapunov exponents $\lambda_1, \lambda_2$ for each value of $\beta$, characterizing the stability of the system’s dynamics. Positive values of $\lambda_1$ signal chaos and coincide with a phase transition towards large entropy production, while values at zero captures limit cycles.  Together, these results demonstrate how chaotic dynamics lie at the onset of nonequilibrium dynamics with large entropy production.

\section{Discussion}

The results presented in this work demonstrate that asymmetric associative memories offer a compelling framework for constructing interpretable, nonequilibrium CTRNNs capable of rich temporal dynamics. 
Importantly, asymmetric couplings imply that the system no longer derives from a potential and thus escapes the constraints of gradient dynamics. Instead, the network can exhibit nonconservative forces and persistent currents in phase space. This opens a door to robust encoding of temporal structure: sequential activation of patterns, limit cycles, and more complex dynamic attractors, including chaos. Such temporal behaviours are inaccessible to classical associative memory models but emerge naturally in the nonequilibrium regime.

Viewed in this light, asymmetric associative memories serve as interpretable CTRNNs with built-in memory architecture. Pattern vectors $\bm \xi_i$ define a low-dimensional codebook over which trajectories unfold, and the coupling matrix $\bm A$ orchestrates transitions between stored patterns. This perspective highlights how sequence generation, context-sensitive recall, and structured temporal dynamics can all arise from the interplay of asymmetric connectivity and stochastic update rules---and yet remaining analytically tractable.


Future work may explore more thoroughly the capability of such systems to encode more complex configurations of sequential patterns, as well as the learning dynamics that give rise to such nonequilibrium dissipative structures. Furthermore, the work presented here could be studied in implementation in biological models or embodied agents, exploiting the natural connection between associative memories with statistical inference and machine learning setups.

\section{Acknowledgements}
M.A. is partly supported by John Templeton Foundation (grant 62828) and Grant PID2023-146869NA-I00 funded by MICIU/AEI/10.13039/501100011033 and cofunded by the European Union. M.A. and I.G. are supported by the Basque Government through the BERC 2022-2025 program and by the Spanish State Research Agency through BCAM Severo Ochoa excellence accreditation CEX2021-01142-S funded by MICIU/AEI/10.13039/501100011033. D.D.M. and D.S. acknowledge financial support from the grant PIBA 2024 1 0016 (Basque Government). D.D.M. acknowledges financial support from the grant Project PID2023-146408NB- I00 funded by MICIU/AEI/10.13039/501100011033 and by FEDER, UE. D.S. would like to acknowledge the DIPC Supercomputing Center for computational resources and their technical support
and assistance.

\footnotesize
\bibliographystyle{apalike}
\bibliography{references} 

\end{document}